\documentclass[12pt,preprint]{aastex}

\newcommand{\chandra}{{\it Chandra}}
\newcommand{\xmm}{{\it XMM-Newton}}
\newcommand{\conx}{{\it Constellation-X}}
\newcommand{\ee}[1]{\times10^{#1}}
\newcommand{\dem}{\mathcal{D}}
\newcommand{\ddem}{\mathcal{H}}

\newcommand{\nH}{n_{\rm H}}
\newcommand{\kms}{km\,s$^{-1}$}

\slugcomment{{\Large DRAFT: \today}}

\begin{document}

\title{Ion-by-Ion Differential Emission Measure Determination of
Collisionally Ionized Plasma: I. Method}

\shorttitle{Ion-by-Ion DEM Determination: I. Method}

\author{Patrick S. Wojdowski and Norbert S. Schulz}

\affil{Center for Space Research, Massachusetts Institute of
Technology}

\email{pswoj@space.mit.edu}

\begin{abstract}
We describe a technique to derive constraints on the differential
emission measure (DEM) distribution, a measure of the temperature
distribution, of collisionally ionized hot plasmas from their X-ray
emission line spectra.  This technique involves fitting spectra using
a number of components, each of which is the entire X-ray line
emission spectrum for a single ion.  It is applicable to
high-resolution X-ray spectra of any collisionally ionized plasma and
particularly useful for spectra in which the emission lines are
broadened and blended such as those of the winds of hot stars.  This
method does not require that any explicit assumptions about the form
of the DEM distribution be made and is easily automated.
\end{abstract}

\section{Introduction}
\label{sec:intro}
The X-ray emission for a large class of astrophysical objects is
dominated by emission from a hot, thin plasma in which the electrons
have a thermal velocity distribution and ion-electron recombination is
balanced by collisional ionization --- i.e., collisional ionization
equilibrium.  This class of astrophysical objects includes both hot
and cool stars, clusters of galaxies, and elliptical galaxies as well
as some cataclysmic variable systems and active galactic nuclei
\citep[see, e.g.,][]{pae03}.  A large class of laboratory plasmas
%including electric discharges, tokamak plasmas, laser-produced plasmas
are also in collisional ionization equilibrium.  At temperatures above
$\sim10^6$\,K, plasmas in collisional ionization equilibrium emit
X-rays in the 1.5--30\,\AA{} (0.4--8\,keV) band.  Up to temperatures of
$\sim3\ee8$, that X-ray emission is characterized by strong line
emission.  This line emission varies strongly with temperature changes
of a few tenths of a decade or more.

The unprecedented ability of the grating spectrometers on the
\chandra{} and \xmm{} observatories to resolve this X-ray line
emission present new possibilities for measuring temperatures of
plasmas in CIE and, therefore, new tests of theories of astrophysical
objects.  Compilations of X-ray line emissivities have long been
available (e.g., \citealt{ray77}, \citealt*{mew85}, \citealt{smi01})
and given a detailed model of an astrophysical object, it is, in
general, straightforward to compute the emission line spectrum of any
optically thin CIE plasma.  For the purpose of determining
luminosities of most emission lines, the three dimensional temperature
and density distributions may be reduced to the differential emission
measure (DEM) distribution, a single-valued function of temperature
which we define in \S\ref{sec:dem}.  However, it is often the case
that no explicit theoretical prediction exists for the DEM
distribution of objects of interest.  Furthermore, even when theory
does provide an explicit prediction of the differential emission
measure, it is often desirable to make measurements of the
differential emission measure without respect to any model.
Therefore, in this paper, we develop a method for obtaining
constraints on the DEM distribution of a collisionally ionized plasma
from its X-ray emission line spectrum without respect to any physical
model.

Our method consists of fitting the observed spectral data using a
model consisting of a continuum plus a number of line emission
components, with each line emission component containing all of the
lines of a single ion in the observed wavelength band.  Since each ion
emits only in a specific temperature range, the best fit magnitude for
the line emission of a given ion gives a measure of a weighted average
of the DEM distribution in that temperature range times the abundance
of that element.  We plot the constraints for all of the ions as a
function of the temperature at which emission from the ion peaks.
These plots may be understood as one-dimensional ``images'' of the DEM
distributions in that they consist of discrete ``pixels'' (one for
each ion) and differ from the true DEM distribution by a convolution
which may be understood as a ``temperature-spread'' function.  In
addition, by fitting the entire X-ray spectrum, rather than attempting
to measure fluxes of individual lines, we take advantage of the
information available from blended lines.  This is particularly
important in the analysis of spectra in which the lines are
significantly broadened and blended due to insufficient instrumental
spectral resolution, Doppler motion of the emitting plasma, or other
effects.  This method can be automated, facilitating the analysis of
large numbers of spectra.  This attribute will become increasingly
important with the future \conx{} mission which will be able to obtain
high-resolution spectra for a large number of objects.  While we are
unaware of any use of this method exactly as it is described here, it
is quite similar to and, in fact, inspired by methods described and
applied by \citet{pot63} for the analysis of solar ultraviolet
spectra, by \citet{sak99} for the analysis of an X-ray spectrum of the
photoionized wind of the high mass X-ray binary Vela~X-1, and by
\citet{bri01} and \citet*{beh01} for analyses of the X-ray spectra of
the corona of the cool stars HR~1099 and Capella, respectively.

In \S\ref{sec:dem} we motivate and describe our method in detail and
in \S\ref{sec:simtest}, we apply it to a simulated spectrum of a
plasma with a continuous temperature distribution and simulated
spectra of several single temperature plasmas.  In
\S\ref{sec:discuss}, we discuss the possible applications of our
method.  The X-ray emission of most hot stars originates in supersonic
stellar winds, resulting in broadened and blended emission lines.
Therefore, our technique is particularly useful in the analysis of
these spectra and in a companion paper (hereafter, ``Paper II'') we
apply our technique to spectra of nine hot stars.

\section{The Differential Emission Measure Distribution and its
Determination}
\label{sec:dem}

In a plasma in CIE, line emission is due primarily to electron-ion
collisional excitation.  Therefore, at low densities, for a given
temperature, line emissivities are proportional to the square of the
density.  However, at high enough density, ions in some metastable
excited states may undergo additional collisional excitation,
resulting in the emissivities of some lines depending on higher powers
of the density.  By definition, in a plasma in CIE, the radiation
field is too weak to affect the ionization balance.  However, there is
a regime in which a plasma may be in CIE but in which the radiation
field is strong enough to induce further excitation of ions in
metastable excited states.  This leads to additional dependence of the
line emissivities on the radiation intensity.

For simplicity, in \S\ref{sec:low_den}, we define the DEM distribution
and describe a method for deriving constraints on it from its X-ray
emission spectrum for a plasma in which excited ions do not undergo
further excitation.  Then, in \S\ref{sec:pump}, we describe a modified
version of this method for deriving constraints on the DEM
distribution of plasmas in which ions in metastable excited may
undergo further excitation.

\subsection{Low Density and Radiation Intensity}
\label{sec:low_den}

In the limit of low density and low radiation intensity, at a given
temperature, line emissivities depend on the square of the density.
Therefore, for the purpose of determining the emission line spectrum,
the three dimensional temperature and density distributions may be
reduced to the DEM distribution defined as:
\begin{equation}
\dem(T)\equiv\frac{dE}{d\log{}T}
\label{eqn:dem}
\end{equation}
where $T$ is the electron temperature and $E$ is the emission measure
defined as
\begin{equation}
E(T)\equiv\int_0^{T} n_e\nH{}dV
\end{equation}
where $n_e$ is the electron density, $\nH{}$ is the hydrogen atom
density and the integration is over that volume where the temperature
is less than $T$.  In addition, because diffuse plasmas in collisional
ionization equilibrium cool radiatively, the cooling of gas is also
determined by the DEM distribution.  Therefore, constraints on the DEM
distribution provide constraints on the overall energetics of the
X-ray emitting plasmas.

The luminosity of an emission line $i$ from an ion of charge state $z$
of element $Z$ may be expressed as
\begin{equation}
L_{Z,z,i}=A_Z\int \dem(T)P_{Z,z,i}(T)d\log{}T
\label{eqn:l}
\end{equation}
where $A_Z$ is the abundance of element $Z$ relative to solar and
$P_{Z,z,i}(T)$ is the line power function.  The line
power function depends only on atomic physics parameters and the solar
abundance of the element.  Its variation with temperature is due to
the variation of the fractional abundance of the emitting ion as well
as the variation of intrinsic collisional excitation rates.  Line
power functions have been tabulated for a large number of lines.
Therefore, from the measurement of a line luminosity, a constraint on
the product of the elemental abundance and the DEM distribution can be
inferred.

\begin{figure}
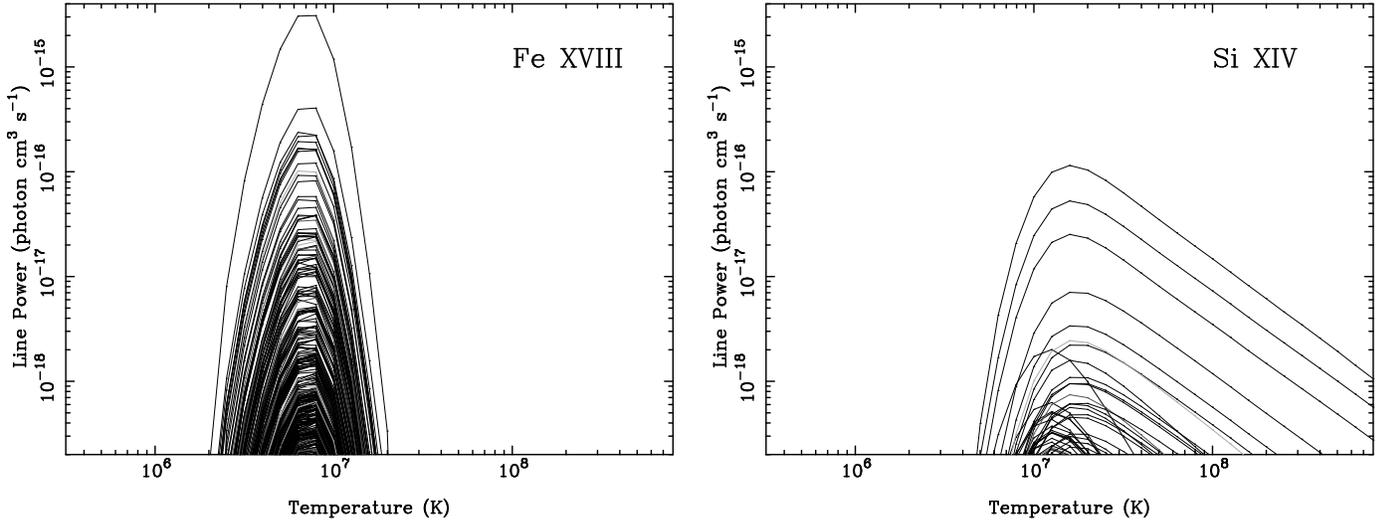

\begin{tabular}{cc}
\includegraphics[angle=-90,width=3.5in]{f1a.eps} &
\includegraphics[angle=-90,width=3.5in]{f1b.eps} \\
\end{tabular}
\caption{Line power functions and summed line power functions for the
$L$-shell ion \ion{Fe}{18} and the hydrogen-like ion \ion{Si}{14}.
The topmost curve in each panel is the total line power function
($\Theta$, defined by Equation~\ref{eqn:Theta}).  All lines from the
database are plotted.  For \ion{Fe}{18}, the lines can be seen to have
nearly the same temperature dependence.  For \ion{Si}{14}, a family of
line power functions have a temperature dependence significantly
different from the others, peaking at a lower temperature.  However,
the peak values of these line power functions are only a few percent
of those of the strongest lines.  These figures illustrate
the validity of our approximation that the line power functions of
most of the lines of a given ion, and all of the strongest lines of
that ion, the line power functions have almost the same shape.  The
sparseness of the temperature grid is evident from the jagged
appearance of the curves.}
\label{fig:t_deps}
\end{figure}

In the absence of specific theoretical predictions of DEM
distributions, it may be useful to determine, at least approximately,
the magnitude and form of the DEM distribution.  If we divide
Equation~\ref{eqn:l} by $\int{}P_{Z,z,i}(T)d\log{}T$, then it may be
seen that the measurement of a line luminosity constrains the product
of an elemental abundance and a ``weighted average'' of the DEM
distribution.  In Figure~\ref{fig:t_deps}, we show the power functions
of the lines of two ions.  
%%% START FIGURE
%%% END FIGURE
The line power data we use in this figure and elsewhere in this paper
is from a version of Astrophysical Plasma Emission Database
\citep[APED,][]{smi01} based on the publicly released version 1.1 but
modified to include the dependence of the line power functions on
density (R. Smith, private communication 2002).  The grid for the
database we use is somewhat sparse: line powers are tabulated for ten
temperature values per decade and two density values per decade.  It
may be seen in the figure that the line power functions are
single-peaked and, compared to their peak values, are negligible
outside of temperature ranges of about one decade around each peak.
Therefore, the weighted average values of the DEM that we obtain from
the measurement of line luminosities are, approximately, the average
value of the DEM in the vicinity of the peak of that line's power
function.

Measuring the luminosities of several lines with power functions that
peak at different temperatures gives us approximate average values of
the DEM distribution in several temperature ranges.  However, for many
applications, it may be difficult to measure luminosities of
individual lines as they may be blended owing either to the finite
spectral resolution of the instrument or intrinsic line broadening.
In order to avoid the uncertainties related to the measurement of
fluxes of individual lines, we take advantage of the fact that the
power functions of most of the lines, and all of the strong lines, of
an ion have very nearly the same shape.  This similarity of line power
functions is demonstrated in Figure~\ref{fig:t_deps}.  If two lines
have power functions that have the same shape but differ by a constant
factor, then the ratio of the luminosities of those two lines will
differ by that same constant factor regardless of the DEM
distribution.  Therefore, instead of attempting to measure the fluxes
of individual emission lines in a spectrum, we fit the entire spectrum
using all of the lines in the database, constraining the line fluxes
such that ratios of the line fluxes for any individual ion are fixed.
In such a fit, there is only one free line luminosity normalization
parameter for each ion.  Just as the luminosities of individual lines
imply average values of the DEM distribution in temperature ranges
defined by the line power functions, so do these normalization
parameters.  However, because each ion has a distinct set of lines
with distinct luminosity ratios, these normalization parameters and
the corresponding average values of the DEM distribution can be
uniquely constrained, even if lines are significantly blended.  For
several ions, there is a family of lines, the dielectronic
recombination lines, that have power functions that differ
significantly from the other lines of the ion.  This fact can be seen
in Figure~\ref{fig:t_deps} for \ion{Si}{14}.
%Furthermore, even neglecting the dielectronic recombination lines,
%the line power functions do not have exactly the same shape.  
However, the dielectronic lines are quite weak.  The peak powers of
the dielectronic recombination lines are only a few percent of those
of strongest lines.  Therefore, we proceed with our attempt to derive
constraints on DEM distributions by fitting spectra with fixed line
luminosity ratios.

In order to choose line luminosity ratios for fitting spectra, we
define the function $\Theta_{Z,z}$ for each ion as the sum of all of
the power functions for all of the lines of that ion:
\begin{equation}
\Theta_{Z,z}(T)\equiv\sum_iP_{Z,z,i}(T).
\label{eqn:Theta}
\end{equation}
For each function $\Theta$, we define a temperature $T_{\rm p}$ to be
the temperature at which it peaks.  
We then let the line luminosities be
given by
\begin{equation}
L_{Z,z,i}=D_{Z,z}P_{Z,z,i}(T_{{\rm p},Z,z})\Delta\log{}T_{Z,z}
\label{eqn:ld}
\end{equation}
where $D_{Z,z}$ is a variable normalization parameter and
\begin{equation}
\Delta_{Z,z}\log{}T\equiv\frac{\int\Theta_{Z,z}(T)d\log{}T}{\Theta_{Z,z}(T_{{\rm p},Z,z})}.
\label{eqn:dellogt}
\end{equation}
In Table~\ref{tab:ion_temp} we give values of $T_{{\rm p}}$ and the
quantities $\Delta_{-}\log{}T$ and $\Delta_{-}\log{}T$, defined as 
\begin{eqnarray}
\Delta_{-}\log{}T_{Z,z} & \equiv & \frac{\int_0^{T_{{\rm p},Z,z}}
\Theta_{Z,z}(T)d\log{}T}{\Theta_{Z,z}(T_{{\rm p},Z,z})},\\
\Delta_{+}\log{}T_{Z,z} & \equiv & \frac{\int_{T_{{\rm
p},Z,z}}^\infty \Theta_{Z,z}(T)d\log{}T}{\Theta_{Z,z}(T_{{\rm p},Z,z})}
\end{eqnarray}
 for each of
the ions we use in our analysis.  The quantities $\Delta_{-}\log{}T$ and
$\Delta_{+}\log{}T$ provide an indication of the temperature ranges
over which the ions have significant line emission. 
\begin{deluxetable}{lrrr}
\tablewidth{0pt}
\tablecaption{Temperature Data for Ion Line Power
Functions\label{tab:ion_temp}}
\tablehead{
\colhead{Ion} & \colhead{$T_{\rm p}$(K)} &
%\colhead{$\Delta\log{}T$} & 
\colhead{$\Delta_{-}\log{}T$} &
\colhead{$\Delta_{+}\log{}T$}}
\startdata
\ion{N}{6}   & $1.6\ee6$ & 0.20 & 0.34 \\
\ion{N}{7}   & $2.0\ee6$ & 0.14 & 0.34 \\
\ion{O}{7}   & $2.0\ee6$ & 0.16 & 0.21 \\
\ion{O}{8}   & $3.2\ee6$ & 0.18 & 0.34 \\
\ion{Ne}{9}  & $4.0\ee6$ & 0.22 & 0.19 \\
\ion{Ne}{10} & $6.3\ee6$ & 0.23 & 0.35 \\
\ion{Mg}{11} & $6.3\ee6$ & 0.22 & 0.22 \\
\ion{Mg}{12} & $1.0\ee7$ & 0.20 & 0.40 \\
\ion{Si}{13} & $1.0\ee7$ & 0.24 & 0.23 \\
\ion{Si}{14} & $1.6\ee7$ & 0.21 & 0.42 \\
\ion{S}{15}  & $1.6\ee7$ & 0.30 & 0.22 \\
\ion{S}{16}  & $2.5\ee7$ & 0.24 & 0.43 \\
\ion{Ar}{17} & $2.0\ee7$ & 0.25 & 0.28 \\
\ion{Ar}{18} & $4.0\ee7$ & 0.28 & 0.42 \\
\ion{Ca}{19} & $2.5\ee7$ & 0.24 & 0.33 \\
\ion{Ca}{20} & $5.0\ee7$ & 0.23 & 0.48 \\
\ion{Fe}{17} & $5.0\ee6$ & 0.18 & 0.21 \\
\ion{Fe}{18} & $7.9\ee6$ & 0.21 & 0.10 \\
\ion{Fe}{19} & $7.9\ee6$ & 0.11 & 0.15 \\
\ion{Fe}{20} & $1.0\ee7$ & 0.13 & 0.11 \\
\ion{Fe}{21} & $1.0\ee7$ & 0.08 & 0.18 \\
\ion{Fe}{22} & $1.3\ee7$ & 0.11 & 0.15 \\
\ion{Fe}{23} & $1.6\ee7$ & 0.16 & 0.17 \\
\ion{Fe}{24} & $2.0\ee7$ & 0.18 & 0.32 \\
\ion{Fe}{25} & $6.3\ee7$ & 0.33 & 0.33 \\
\ion{Fe}{26} & $1.3\ee8$ & 0.28 & 0.48 
\enddata
\end{deluxetable}
%We will make use of the quantities $\Delta_{-}\log{}T$ and
%$\Delta_{+}\log{}T$ in \S\ref{sec:implementation}, however
The fact that several ions have the same value of $T_{\rm p}$ is an
artifact of the sparse temperature grid.

%The data base we use has line powers as functions of density in the
%range $10^6$--$10^{15}$\,cm$^{-3}$ and of temperature in the range 
%$10^4$--$7.9\ee8$\,K.  
%In all computations described in this paper,
%line power functions are taken to be zero outside that temperature
%range. 
% We first determine $T_{{\rm p},Z,z}$ and $\Delta_{Z,z}\log{}T$
%for each ion using $n_e=10^6$\,cm$^{-3}$, the lowest density in the
%database.  

If all of the line power functions of a given ion have exactly the
same shape, then the value of $D$ for that ion gives the product of
the elemental abundance and the average of the DEM distribution
weighted by that ion's function $\theta$ as described above for
individual lines.  As mentioned above, not all of the line power
functions for every ion have the same shape.  However, as those line
power functions that differ significantly have magnitudes of only
approximately 1\% of the value of the strongest line power functions,
we expect an error due to the difference of line power functions of no
more than about 1\%.

\subsection{Line Pumping}
\label{sec:pump}

%We are not interested here in diagnosing the density distribution of
%the X-ray emitting material or the intensity of the radiation field in
%it.  However, the X-ray emission line spectra of hot stars indicate
%significant radiation densities in the emitting plasma.  Therefore, a
%small modification to the method described in \S\ref{sec:low_den} is
%required to fit these spectra and obtain accurate constraints on their
%DEM distributions.

As we have mentioned before, some ions have metastable excited states
that are susceptible to further excitation by collisions with
electrons or absorption of photons.  The result of this is that the
emission in lines resulting from the decay of the metastable state are
``pumped'' into other lines.  We first consider pumping by collisional
excitation because the effects of the two mechanisms are similar and
because data for pumping by collisional excitation is more readily
available.  For this case, the line emission may still be described by
line power functions.  However, the line power $P$ is a function of
density as well as temperature.  Therefore, the emission line spectrum
depends on the two-dimensional temperature-density DEM
\begin{equation}
\ddem(T,n_e)\equiv{}\frac{d^2E}{d\log{}T\,d\log{}n_e}
\end{equation}
The DEM distribution is related to this quantity by
\begin{equation}
\dem(T)=\int\ddem{}(T,n_e)d\log{}n_e.
\end{equation}
If the density dependence of each line power function of a single ion
is due to a single upward transition from a single metastable state
and the collisional excitation rate for that transition does not
change much over the temperature range where that ion emits, then the
line power functions for that ion may be well approximated by the form
\begin{equation}
P_{Z,z,i}(T,n_e)=\Theta_{Z,z}(T)(B_{Z,z,i}+F_{Z,z,i}H_{Z,z}(n_e))
\label{eqn:ptd}
\end{equation}
where $\Theta$ and $H$ are functions that are the
same for all of the lines of a given ion and $B$ and
$F$ are constant coefficients for each line.  Even if the two
conditions mentioned above are not satisfied, the line power functions
may be well approximated by this form.  We discuss the validity of
this assumption in \S\ref{sec:accpow} and proceed here with the
assumption that the power functions have this form.

We define the functions $\Theta$, and the temperatures where they
peak, $T_{\rm p}$, as before in the limit of low density and low
radiation intensity.  With this definition, the functions $H(n_e)$ go
to zero as $n_e$ does.  If the functions $H$ are continuous, it is
possible to show that for any temperature-density DEM distribution
$\ddem(T,n_e)$ with an associated DEM distribution
$\dem(T)=\int{}\ddem(T,n_e)d\log{}n_e$, there exists, for each ion, a
value of the density $n^\prime_{Z,z}$ such that the
temperature-density DEM distribution $\ddem_{Z,z}^\prime$ defined as
\begin{equation}
\ddem_{Z,z}^\prime(T,n_e)\equiv\dem(T)\delta(n_e-n^\prime_{Z,z})
\end{equation}
produces the same luminosities for all of the emission lines of the
ion $Z,z$ as does $\ddem(T,n_e)$.  Therefore, we modify
the method described above by adopting the variable parameters
$n^\prime_{Z,z}$ and let the line luminosities be
\begin{equation}
L_{Z,z,i}=D_{Z,z}P_{Z,z,i}(T_{{\rm p},Z,z},n^\prime_{Z,z})\Delta_{Z,z}\log{}T
\end{equation}
where $\Delta_{Z,z}\log{}T$ is defined as before in
Equation~\ref{eqn:dellogt}.  That is, we fit the entire spectrum using
all of the lines in the database as described in \S\ref{sec:low_den}.
However, instead of constraining the line flux ratios of individual
ions to be fixed, we allow the ratios to vary with a density parameter
$n^\prime_{Z,z}$ for that ion.  The density parameters for each of the
ions are allowed vary independently.  Again, $D_{Z,z}$ approximates
the product of the elemental abundance and the average of the DEM
distribution weighted by $\Theta_{Z,z}(T)$.  Even if
equation~\ref{eqn:ptd} is not satisfied, this method will result in
good fits to the data and accurate constraints on the DEM distribution
if the emitting plasma has a single density or if density and
temperature are strongly correlated within the plasma.

\subsubsection{Accuracy of the Power Function Approximation}
\label{sec:accpow}

A systematic study of the density dependence of emission lines with
wavelengths from 1.2--31\,\AA{} has been undertaken by \citet{smi02}.
These authors have made fits to the line powers as functions of
density at the constant temperatures $10^6$, $10^{6.5}$, $10^7$, and
$10^{7.5}$~K for lines with peak powers exceeding a minimum value and
also meeting a criterion for variability with density at each
temperature.
%These authors also searched for lines meeting these criteria for the
%temperatures $10^5$, $10^{5.5}$, and $10^8$\,K but found none. 
They found the power functions of most of the lines satisfying those
criteria ($\sim$90\%) could be approximated well using a function of
the form
\begin{equation}
P(n_e)=c_0+c_1\exp(-n_e/n_1)
\label{eqn:pd}
\end{equation}
where $c_0$, $c_1$, and $n_1$ are fit parameters.  If the density
dependence for the line powers of all of the lines of an ion have the
form of Equation~\ref{eqn:pd}, then Equation~\ref{eqn:ptd} is a valid
description of the line power functions if, in the temperature range
where line emission is significant,
\begin{itemize}
\item all of the lines of any one ion have the same value of $n_1$ and
\item for any single line, the ratio of the line power at high density
($c_0$) to the value at low density ($c_0+c_1$) and the value of $n_1$
do not change with temperature.
\end{itemize}
We have inspected the results of \citet{smi02} and found that, for all
of the ions we use except \ion{Fe}{19}, \ion{Fe}{20}, and
\ion{Fe}{21}, most of the lines ($\sim$90\%) are described by
Equation~\ref{eqn:pd} and have values of $n_1$ that are very close,
having a standard deviation of 0.1 or less in $\log{}n_1$.  For those
lines, of any ion, that meet the criteria to be fit at more than one
temperature, the root mean square (RMS) of the difference between
values of $\log{}n_1$ for the same ion at different temperatures is
0.09 and the RMS of the difference between values of
$\log(c_0/(c_0+c_1))$ is 0.16.  In summary, Equation~\ref{eqn:ptd} is
not exactly satisfied for all ions.  However, because most line power
functions do not depend on density, it is unlikely that this would
cause errors greater than a factor of a few.  Because
Equation~\ref{eqn:ptd} is near to being satisfied for most ions, we
expect the actual errors to be much less: not much greater than 10 or
20\%.

\subsubsection{Radiation}

Because metastable states may be photoexcited, line powers are
functions not only of temperature and density but also of the mean
radiation intensity at the frequencies of the transitions that affect
line emission.  That is, the line power may be written
$P_{Z,z,i}(T,n_e,J_\nu)$ or
$P_{Z,z,i}(T,n_e,J_{\nu_{i1}},J_{\nu_{i2}},...,J_{\nu_{im}})$ where
$J_\nu$ is the mean radiation intensity as a function of frequency and
$J_{\nu_{ij}}$ are the mean radiation intensities at the frequencies
of the transitions affecting line $i$.  While this is, in principle, a
large number of variables, the number of transitions in which
photoexcitation plays a significant role (the value of $m$) is often
only one.  Furthermore, for a given system, the values of
$J_{\nu_{ij}}$ may be a function of a small number of variables.  For
example, in hot star winds, the radiation intensity due to the stellar
photosphere is given by
\begin{equation}
J_\nu=I_{\star,\nu}W
\end{equation}
where $I_{\star,\nu}$ is the radiation intensity at the surface of the
star and is approximately that of a blackbody with a temperature,
depending on the stellar type, of a few tens of thousand K and $W$ is
a factor accounting for the geometrical dilution of the stellar
radiation with distance from the star.  We know of no systematic study
(at least, not of the scope of that by \citealt{smi02} for density) of
the dependence of line powers on radiation.  However, as radiation and
density both affect the line power through excitation of metastable
stable states, the effects are similar.  Therefore, we use the method
described above --- we take the line emission from each ion to be
determined by the temperature at which its emission peaks and a
density which is a free parameter, independent of the density values
of the other ions --- and assume that the effects of radiation and
density can be replicated approximately by density alone.  In Paper II
we again address the validity of this approach for plasmas with a
significant pumping radiation field.

\section{Tests with Simulated Data}
\label{sec:simtest}

\subsection{Simulations}
\label{sec:sims}

In order to demonstrate the performance of our model, we tested our
method by applying it to a series of simulated spectra.  We simulated
\chandra{} MEG and HEG spectra for a plasma with a DEM distribution
constant with temperature over the range
$3.2\ee{5}$--$7.9\ee8$\,K\footnote{actually one component for each of
the database's temperature values, with each component having the same
emission measure} and zero elsewhere.  We also simulated spectra from
single-temperature plasmas at the temperatures $3\ee5$, $10^6$,
$3\ee6$, $10^7$, $3\ee7$, $10^8$, and $3\ee8$\,K.  In Paper II we
apply our method to spectra of hot stars.  Therefore, for our
simulations we adopt the parameters of $\zeta$~Pup, the best-studied
hot star in the X-ray band, and of an observation of it with
\chandra{} (see Paper II).  However, the results of the application of
our method to these simulated data indicate the general behavior of
our method.

To choose a total emission measure for our simulations, we conducted a
number of three-temperature fits to the spectrum of $\zeta$~Pup, using
an fixed value of $1.0\ee{22}$\,cm$^{-2}$ \citep*[c.f.,][]{ber96} for
the interstellar absorption column.  These fits resulted in total
emission measures in the range (2.4--5.0)$\ee{55}$\,cm$^{-3}$ (for an
adopted distance of 450\,kpc, \citealt*{sch97}) depending on whether
we tried to fit the nitrogen lines or the oxygen lines (this
discrepancy is discussed in more detail in Paper II) and we
chose a value of $3.6\ee{55}$\,cm$^{-3}$ for the total emission
measure.  For the constant DEM distribution, this implies a DEM value
of $1.0\ee{55}$\,cm$^{-3}$.  For our simulated exposure time, we used
68,598\,s, the exposure time of our observation of $\zeta$~Pup.  For
all of the simulations, $n_e$ was taken to be
$8\times10^{13}$\,cm$^{-3}$.  We chose this particular density value
because it is large enough so that, like in the actual stellar
spectra, the forbidden lines of the helium-like triplets are
completely pumped into the intercombination lines.  For the simulated
line profiles we used the ISIS thermal/turbulent line profile
function:
\begin{equation}
\phi_{Z,z,i}(\lambda)=\frac{1}{\sigma_{Z,z}\lambda_{Z,z,i}\sqrt{2\pi}}
\exp\left(\frac{(\lambda/\lambda_{Z,z,i}-(1+v_{\rm r}/c))^2}
{2\sigma_{Z,z}^2}\right)
\label{eqn:l_profile}
\end{equation}
where
\begin{equation}
\sigma_{Z,z}\equiv{}c^{-1}
\left(\frac12v_{\rm t}^2+\frac{kT_{{\rm p},Z,z}}{m_Z}\right)^{1/2}
\end{equation}
where $m_Z$ is the mass of element $Z$, $k$ is Boltzmann's constant.
We took $v_{\rm r}$ to be 0 and $v_{\rm t}$ to be 800\,\kms,
approximating the line widths observed by \citet{kah01} for
$\zeta$~Pup.

\subsection{Application of the Method to the Simulated Data}
\label{sec:implementation}

We fit the simulated spectra using the method described in
\S\ref{sec:pump}.  In our analysis we do not attempt to use
measurements of the continuum to constrain the DEM distribution.
However, in order to fit spectra and obtain accurate constraints from
the emission lines, it is necessary to account for the continuum.  The
continuum emission from a collisional plasma is due primarily to
bremsstrahlung, though radiative recombination continua and two-photon
continua also contribute.  For our fits, we use a continuum consisting
of three bremsstrahlung components with variable temperatures and
normalizations.  In all cases, this provides a sufficient empirical
representation of the continuum.  For the fit model line profiles, we
use the same function as for the simulation model line profile
(equation~\ref{eqn:l_profile}).  However, in the fit, the values of
$v_{\rm r}$ and $v_{\rm t}$ are taken to free but to have the same
values for all lines.  Also, in the fit spectral models, we included
the same absorption ($1.0\ee{22}$\,cm$^{-2}$) used for the
simulation.  

Though we use the same atomic database for the simulations and the
model we fit to the simulated data, this does not amount to fitting
the data with the same model used for the simulation.  For the
simulations, the line luminosities are given by equation~\ref{eqn:l}
with $\dem(T)$ constant or proportioal to $\delta(T-T_{\rm sim})$
where $T_{\rm sim}$ is the temperature of the simulated plasma.  On
the other hand, in the model fit to the simulated data, the line
luminosites are given by equation~\ref{eqn:ld}.  Unless all of line
power functions have the same shape every other line power function of
the same ion, the line luminosities of the simulation model and the
fit model will necessarily differ.  Therefore, the fits to the
simulated data test whether or not the differences in line power
shapes between lines of individual ions are small enough for our
method to be valid.

%  The decision
%as to how many bremsstrahlung components are necessary to obtain a
%good fit to the continuum is made somewhat subjectively.  
%However,
%this is the only part of the procedure in which human consideration is
%required for each spectrum.

We simultaneously fit both the HEG and MEG simulated data to find the
best fit values of $D_{Z,z}$ and $n_{Z,z}^\prime$ for each of the ions by
minimizing the \citet{cas79} statistic.  This statistic, unlike the
$\chi^2$ statistic, is valid in the regime where the number of counts
in a channel is small such as is the case for several of our data
sets.  We search for a minimum of the Cash statistic with ISIS, first
by using its implementation of the Levenberg-Marquardt algorithm and
then using its implementation of the simplex algorithm.  We then
search for confidence intervals, again using ISIS, on the parameters
$v_{\rm r}$, $v_{\rm t}$, and on each of the values of $D_{Z,z}$ and
$n_{Z,z}$.  In searching for confidence intervals, only the
Levenberg-Marquardt algorithm is used.  In searching for confidence
intervals, a new minimum is frequently found, requiring the process to
be restarted.  On our workstations (with clock speeds of order
1\,GHz), fitting and finding all of these confidence intervals for one
of our data sets generally takes a few weeks.

\begin{figure}
\plotone{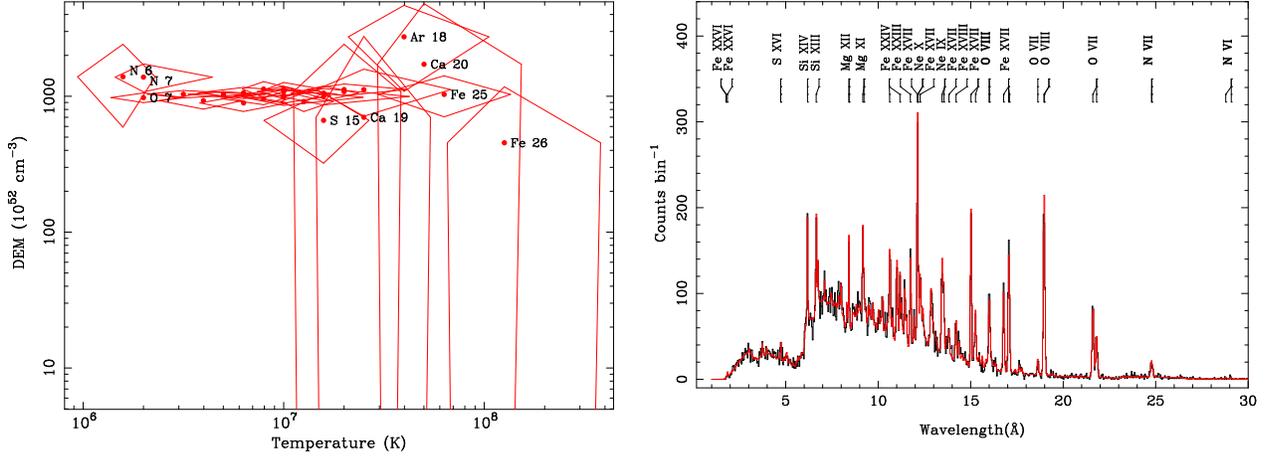}
\caption{In the first panel, the results for the simulated data for a
DEM constant in the range $3.2\ee5$--$7.9\ee8$\,K
($\dem=1.0\ee{55}$\,cm$^{-3}$) is plotted.  For each ion $Z,z$, we
plot a filled circle at ($T_{{\rm p},Z,z}$,$D_{Z,z}$) where $D_{Z,z}$
is our best-fit value.  The vertical extent of the diamond around a
filled circle indicates the confidence region determined for $D_{Z,z}$
and the horizontal extent of a diamond is given by $\Delta_{-}T_{Z,z}$
and $\Delta_{+}T_{Z,z}$, which is defined in the text and is,
approximately, the region over which the ion emits.  The ions
corresponding to the data points are labeled using Arabic numerals
rather than Roman numerals.  Because the data points are so close in
this plot, we label only a few of them.  In the second panel we show
the simulated MEG spectrum (black) and the best-fit model obtained
with our method (red).  We label several bright lines.  Though we do
not show it here, the HEG spectrum was also used in the fit.  For this
and all of our simulated spectra, we adopt the absorption column
($1.0\ee{20}$\,cm$^{-2}$), total emission measure
($3.6\ee{55}$\,cm$^{-3}$) emission measure, and line width (800\,\kms)
of $\zeta$~Pup.  The fact that the data points in the first panel are
consistent with $\dem=1.0\ee{55}$\,cm$^{-3}$ and the residuals are no
larger than the square root of the number of counts indicates that our
method is valid.
%\log{}T_{{\rm p},Z,z}-\Delta_{Z,z,{\rm
%l}}\log{}T$--$\log{}T_{{\rm p},Z,z}$ 
%are the best-fit values of $C_{Z,z}$ for the
%fit to .  As
%expected, these values are, within errors, equal to the value of
%$\dem$.  In the second and third panels are the simulated data (black)
%and best-fit model (red) for the MEG (second panel) and HEG (third
%panel).
} 
\label{fig:flat}
\end{figure}

\begin{figure}
\plotone{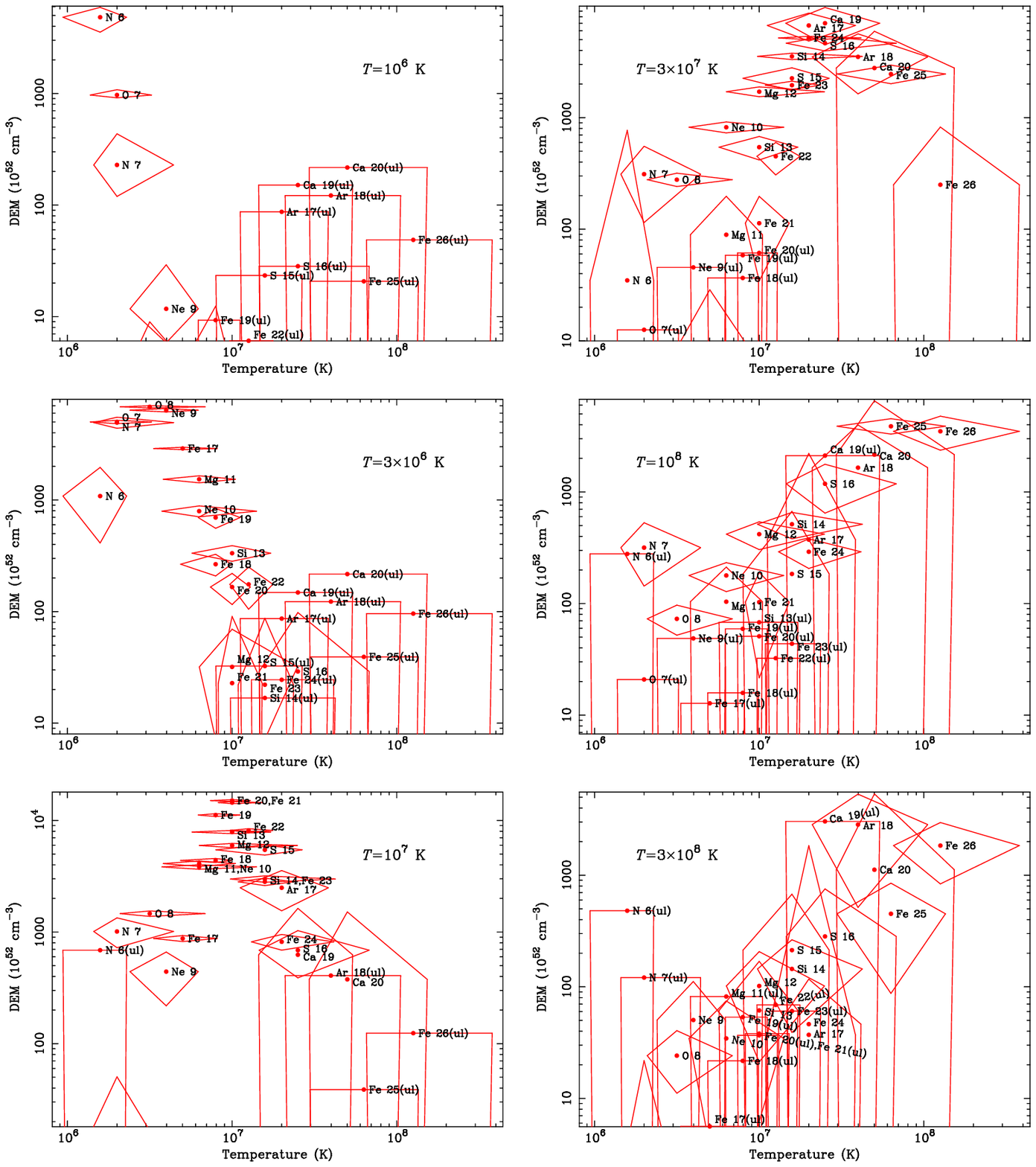}
\caption{Plots of DEM constraints for spectra of plasmas at the single
temperatures $10^6$, $3\ee6$, $10^7$, $3\ee7$, $10^8$, and $3\ee8$ K.
The designation ``(ul)'' indicates an upper limit.}
\label{fig:1temp_dems}
\end{figure}

In Figure~\ref{fig:flat} we plot the best fit values of $D$ as a
function of $T_{\rm p}$ for each of the ions (indicated by filled
circles) and also the simulated data and best fit model spectrum.  In
the plot of $D$ vs. $T_{\rm p}$, a diamond surrounds each filled
circle.  The vertical extent of a diamond indicates the statistical
error on the best-fit value of $D$ (given by $\Delta{}C=2.706$, where
$C$ is the Cash fit statistic --- this is the 90\% confidence region)
and the extent of a diamond to the left and right is given by
$\Delta_{-}\log{}T$ and $\Delta_{+}\log{}T$, respectively.  These
temperature ranges are the temperature ranges for which the values of
$D_{Z,z}$ represent the approximate average values of the DEM.
While we do not assess the quality of the spectral fit quantitatively,
it may be seen that the fit is quite good.  The fact that the data
points in the first panel are consistent with
$\dem=1.0\ee{55}$\,cm$^{-3}$ and the residuals are no larger than the
square root of the number of counts indicates that our method is
valid.  In Figure~\ref{fig:1temp_dems}, we show the plots of $D_{Z,z}$
for the simulated spectra of single-temperature plasmas, except for
the simulated $3\ee5$\,K plasma spectrum which had only a few counts.
The fits for these spectra, which we do not show, are also quite good.
In the first panel of Figure~\ref{fig:flat} and in
Figure~\ref{fig:1temp_dems} we use the same temperature range on the
horizontal axis and 3 orders of magnitude in DEM on the vertical axis.
This allows slopes of lines in the various plots to be compared.
Our best-fit values of $v_{\rm r}$ and $v_{\rm t}$ have uncertainties
of a few tens of \kms{} and are consistent with the input values of
zero and 800\,\kms, respectively.

It may be seen from Equation~\ref{eqn:l} that, for a constant DEM
distribution, we expect the values of $D_{Z,z}$ to be that constant
value of the DEM and, as illustrated in the first panel of
Figure~\ref{fig:flat}, within the errors, this is indeed what we
find.  This, and the good quality of the fits indicates that the fact
that not all of the power functions of every ion have the same
temperature dependence does not cause our method to be significantly
inaccurate.  As our simulated plasma has only a single density, this
does not test whether or not deviations in the line power functions
from the form of Equation~\ref{eqn:ptd} cause significant
inaccuracies in our DEM determinations.  Using plasmas with
distributions of densities would have provided a test of this.
However, this would not have provided a test of the accuracy of our
method for plasmas with strong ambient radiation fields.  Simulating
the emission of a plasma with an ambient radiation field is beyond the
scope of this work.

For each of the single temperature simulations, the determined values
of $D_{Z,z}$ show a peak at the temperature of the simulated plasma.
However, the sharpness of the peak differs for each of the
simulations.  These plots may be understood as ``temperature-spread
functions'' in analogy with point spread functions in images for use
in evaluating our results from observed spectra.

\section{Discussion}
\label{sec:discuss}

We have described a method for deriving constraints on the
differential emission measure distribution of a collisionally ionized
plasma from its X-ray emission spectrum and displaying these
constraints in a way that indicates the DEM distribution.  We have
designed our technique to account for the line pumping that occurs at
high densities and with high radiation intensities.  We have applied
our method to simulated spectra and demonstrated that, with this
method, we can recover simulated DEM distributions in, approximately,
the temperature range $10^6$--$3\ee{8}$\,K from their emission line
spectra subject to a temperature resolution comparable to the width of
the line power functions: a few tenths of a decade.

%We have attempted to design our to work
%In our analysis, we also derive a characteristic line shift
%and width for each spectrum.  We intend to post our analysis scripts to the
%\chandra{} contributed software exchange
%(\url{http://asc.harvard.edu/cont-soft/soft-exchange.html}).
We expect that the technique described here will be useful in the
analysis of high-resolution spectra of plasmas in collisional
ionization equilibrium in several astrophysical contexts.  This
technique is particularly useful in cases where lines are broadened to
the point of being blended.  Therefore, we expect this technique to be
especially useful in the analysis of the X-ray spectra of hot stars
and we apply it to spectra of nine hot stars in Paper II.
\citet{muk03} have shown that the X-ray spectra of some cataclysmic
variable stars are characteristic of plasmas in collisional ionization
equilibrium.  Some of those spectra show lines as broad as
500\,\kms{}.  Therefore, this technique may be useful in the analysis
of those objects.  Even for spectra in which lines are not broad and
blended, this technique has advantages over others.  It does not
require fitting of individual lines to determine their fluxes and is
therefore easily automated.  Also with this technique, it is not
necessary to make assumptions or introduce biases about the form of
the DEM distributions.  Therefore, our technique may also be useful
for analysis of high-resolution spectra of the coronae of cool stars
obtained with \xmm{} and \chandra.  The {\it Astro-E II} observatory
will be able to obtain high-resolution spectra of extended objects.
Therefore, with the data from that observatory, our technique may be
useful in the analysis of the spectra of clusters of galaxies and
halos of elliptical galaxies.  The fact that this algorithm can be
easily automated may be of particular importance when the DEM
distributions need to be determined for a large number of spectra.
The fact that the anticipated {\it Constellation-X} observatory will
have a very large effective area for high-resolution spectroscopy will
enable it to obtain spectra for a large number of objects that are too
faint to be efficiently observed with the three previously mentioned
observatories.  Therefore, our technique may be especially useful in
the analysis of the large number of high-resolution X-ray spectra that
will be obtained with that observatory.

\acknowledgements

We thank John Houck for assistance implementing our analysis technique
in ISIS, Dan Dewey for a careful reading of the manuscript, and the
referee, Ehud Behar, for helpful comments.  Support for this work was
provided by the National Air and Space Administration (NASA) through
Chandra Award Number GO0-1119X by the Chandra X-ray Observatory Center
(CXC) which is operated for and on behalf of NASA by the Smithsonian
Astrophysical Observatory (SAO) under contract NAS8-39073.  Support
for this work was also provided by NASA through contract NAS8-01129
and by the SAO contract SVI-61010 for the CXC.

\bibliographystyle{apj} 
\bibliography{../ion_dem}

\end{document}